# 3D sub-nanoscale imaging of unit cell doubling due to octahedral tilting and cation modulation in strained perovskite thin films

Magnus Nord[1], Andrew Ross[1,2], Damien McGrouther[1], Juri Barthel[3], Magnus Moreau[4], Ingrid Hallsteinsen[4,5], Thomas Tybell[4], Ian MacLaren[1]

1. School of Physics and Astronomy, University of Glasgow, Glasgow G12 8QQ, UK
2. Institute of Physics, Johannes Gutenberg Universität Mainz, Staudingerweg 7, 55129, Mainz, Germany
3. Ernst Ruska-Centre, Forschungszentrum Jülich GmbH, 52425 Jülich, Germany
4. Department of Electronic Systems, NTNU - Norwegian University of Science and Technology, 7491 Trondheim, Norway
5. Advanced Light Source, Lawrence Berkeley National Lab, California 94720, USA

**Abstract**

Determining the 3-dimensional crystallography of a material with sub-nanometre resolution is essential to understanding strain effects in epitaxial thin films. A new scanning transmission electron microscopy imaging technique is demonstrated that visualises the presence and strength of atomic movements leading to a period doubling of the unit cell *along* the beam direction, using the intensity in an extra Laue zone ring in the back focal plane recorded using a pixelated detector method. This method is used together with conventional atomic resolution imaging in the plane perpendicular to the beam direction to gain information about the 3D crystal structure in an epitaxial thin film of $LaFeO_3$ sandwiched between a substrate of (111) $SrTiO_3$ and a top layer of $La_{0.7}Sr_{0.3}MnO_3$. It is found that a hitherto unreported structure of $LaFeO_3$ is formed under the unusual combination of compressive strain and (111) growth, which is triclinic with a periodicity doubling from primitive perovskite along one of the three <110> directions lying in the growth plane. This results from a combination of La-site modulation along the beam direction, and modulation of oxygen positions resulting from octahedral tilting. This transition to the period-doubled cell is suppressed near both the substrate and near the $La_{0.7}Sr_{0.3}MnO_3$ top layer due to the clamping of the octahedral tilting by the absence of tilting in the substrate and due to an incompatible tilt pattern being present in the $La_{0.7}Sr_{0.3}MnO_3$ layer. This work shows a rapid and easy way of scanning for such transitions in thin films or other systems where disorder-order transitions or domain structures may be present and does not require the use of atomic resolution imaging, and could be done on any scanning TEM instrument equipped with a suitable camera.

## I. INTRODUCTION

In complex oxides functional properties are coupled to the crystal structure, hence control of the crystal structure enables engineering of functional properties. An ubiquitous feature of perovskites and related structures is tilting of the oxygen octahedra. Whilst this has been well-known for a long time, and the simple tilt patterns were classified many years ago in the seminal work of Glazer [1], it is still a lively subject for investigation. Recently, there has been a great development in tuning the tilting pattern using composition or strain to obtain emergent properties in thin films [2,3].

Of special interest has been the evolution to tilt patterns at epitaxial interfaces, and subsequent emergence of novel functional properties. For example, the interface between antiferromagnetic

(AF) XFeO$_3$ (X = La, Bi) and the ferromagnetic La$_{0.7}$Sr$_{0.3}$MnO$_3$ [LSMO] has been shown to have an emergent ferromagnetic moment at the interface [4-6]. In these material systems charge transfer to the $d^5$ Fe$^{3+}$ is prohibited, while their bulk tilt patterns of their oxygen octahedra are not directly compatible. LaFeO$_3$ (space group 62, *Pbnm*) [LFO] has an $a^-a^-c^+$ rotation and LSMO (space group 167, $R\bar{3}c$) has an $a^-a^-a^-$ rotation pattern (Glazer notation) [1]. This mismatch must thus be accommodated at the interface, likely through atomic reconstructions.

Classic studies have either used X-ray or neutron diffraction structure refinements for bulk crystals or selected area electron diffraction for more restricted sample areas, including disentangling more complex orderings in domain-structured ceramics [7,8]. Recently, real-space atomic resolution imaging in the transmission electron microscope (TEM) or scanning transmission electron microscope (STEM) have been used to show structural transitions associated with a change in tilt pattern resulting from elastic strain in thin films, such as the earlier work of MacLaren *et al*. [9] where tilting was suppressed in (Pr$_{0.7}$Sr$_{0,3}$)MnO$_3$ by the application of strong in-plane compressive strain. In the last few years, there have been a number of studies examining the effects of interfaces on octahedral tilting using bright field or annular bright field STEM [10-16]. There have also been tools created for automated processing of such images to reveal oxygen positions and octahedral tilting [17,18].

The disadvantage of the real-space methods reported above is that bright field or annular bright field STEM imaging is technically challenging requiring a perfect microscope and corrector alignment, together with a sample that is aligned to the desired zone axis. Hence, large area mapping is challenging due to sample bending and height/thickness variations.

Also, all the real space methods only give information about tilting in the plane perpendicular to the beam direction. However, information about the ordering along the beam direction is lacking. Whilst one could make two TEM-lamellae thinned in perpendicular directions and then reconstruct the 3D structure that way [19], this does require careful matching to ensure that it is the same structure being viewed from the two perpendicular directions.

Atomic resolution tomography has been attempted for 3D information, but it is still challenging and certainly requires a very high dose to the sample. Large electron doses can often alter the oxide, and recent work has shown that beam damage to perovskites happens well before it is obvious in high-resolution images [20]. In some cases, 3D information can also be inferred from the shape of the columns in 2 dimensions [10,18,21]. An important drawback for this strategy is that the method requires both very high quality TEM-samples, an optimized aberration corrected STEM and a high degree of post processing to reduce the effects of scanning distortions [22,23].

An interesting prospect to obtain information parallel to the beam direction is to use scattering into higher order Laue zones. In the early days of high angle annular dark field (HAADF) STEM imaging, it was speculated by Spence, Zuo and Lynch that the Higher Order Laue Zone (HOLZ) rings may add significant contributions to the contrast in some cases, causing significant deviations from a pure atomic number "Z-contrast", and complicating image interpretation [24]. Later it was shown that selecting specific angles for the ADF detector could be used to show changes in periodicity along the beam direction between different columns in a complex structure of sodium cobaltate [25]. In contrast to these earlier works, recently there have been large advances in STEM imaging (aka 4D STEM) due to the advent of fast readout pixelated direct electron detectors [26-32].

In this letter, we use a refined approach to imaging third dimension periodicity using HOLZ ring analysis. By relying on a 4D STEM approach we combine HOLZ ring intensity, separated from the general HAADF contrast, with regular high-resolution ABF STEM imaging to quantify how oxygen octahedral tilting and A-site modulation appears in 3D. We use this to examine the structural modifications occurring in the LFO layer of a LSMO/LFO heterostructure grown on (111) $SrTiO_3$ [6] (STO), and discuss based on the HOLZ measurements of LFO how oxygen octahedral tilting and A-site modulation are both clamped at the interfaces with both STO substrate and LSMO top layer, due to a change in tilt patterns.

## II.   EXPERIMENTAL METHODS

Growth of the heterostructure is reported elsewhere [6,33]. Cross sectional TEM lamellae were prepared using a FEI Helios Nanolab DualBeam FIB, using a standard lift-out technique. To characterize the TEM lamella perpendicular to the electron beam, conventional atomic resolution STEM imaging was performed using a probe and image corrected JEOL ARM200CF operated at 200 kV using a convergence angle of 20.4 mrad. High Angle Annular Dark Field (HAADF) images were acquired with a detector accepting electrons scattered to between 73-200 mrad. Annular Bright Field images were acquired with a detector set to detect electrons scattered to angles between 11.7 and 22.7 mrad. To acquire full 4D STEM datasets, a fast pixelated 2D Medipix3 direct electron detector with a Merlin readout system [Quantum Detectors Ltd., Harwell, UK] integrated into an JEOL ARM200F was utilized, whereby the final dataset consists of a 4D matrix of two spatial dimensions and two reciprocal space dimensions with each element containing a number of electron counts. This was performed at an appropriately small camera length to allow data out to about 170 mrad to be recorded to the detector in each diffraction pattern. Datasets were recorded using a probe size of 2-3 Å and a step size in the scan of 1.8 Å, and the work was done with the sample aligned to a <110> axis of the STO. Density functional theory (DFT) calculations were performed using The Vienna Ab initio Simulation Package (VASP)[34,35] with the PBE-sol functional[36]. The PAW-PBE potentials supplied with VASP for La, Fe and O were used, with a plane wave cutoff energy of 550 eV. In the calculations Hubbard U potentials of 3 eV for Fe 3d and 10 eV for La 4f were employed.

## III.   RESULTS AND DISCUSSION

In figure 1a a high angle annular dark field (HAADF) STEM image is depicted, showing a typical "Z-contrast" whereby the most intense peaks are the heavier La or Sr atomic columns, while the less intense peaks are the Mn, Fe or Ti atomic columns. In addition to the variations in intensity, there are also subtle differences in the crystal structure across the thin films. In figure 1a the most obvious of these is the elongation of the La columns in the LFO layer. This suggests a "zig-zagging" of the La atoms along the beam direction, resulting in "elliptical" atomic columns as seen previously by Azough *et al.* [21]. Fitting 2D Gaussians to every A-site atomic column using the Atomap software [18] yields a map of the ellipticity of the A-cations (La and Sr columns) as shown in Figure 1b, revealing a distinct ellipticity in the LFO, not observed in the STO or the LSMO. This is a similar method to the *Atomic Column Shape Analysis* of Borisevich *et al.* [10,37].

To reveal the oxygen octahedral tilt pattern, annular bright field (ABF) images were acquired simultaneously with the HAADF data. Here, the light elements are also visible, making it possible to image the oxygen atomic columns. This is shown in Figure 1c, which in addition to the A- and B-cations also shows the oxygen columns. Looking closely at the position of the oxygen columns in the [110] direction, there is a clear "zig-zag" pattern in the LFO region. By fitting 2D Gaussians to every atomic column in the ABF image [18], the centre positions of the less intense oxygen columns can be extracted. This is shown in the rightmost image, which visualizes the variations in distance between these oxygen columns in the [100]-direction.

Both ellipticity and oxygen superstructure is expected for the bulk $a^-a^-c^+$ tilting pattern of LFO. As shown, doing analysis using atomic resolution STEM data yields information about the two spatial dimensions perpendicular to the electron beam, and with high quality data, some information about the last dimension, specifically the alternating shifting of atoms perpendicular to the electron beam, can be extracted using the shape of the atomic columns. [37]

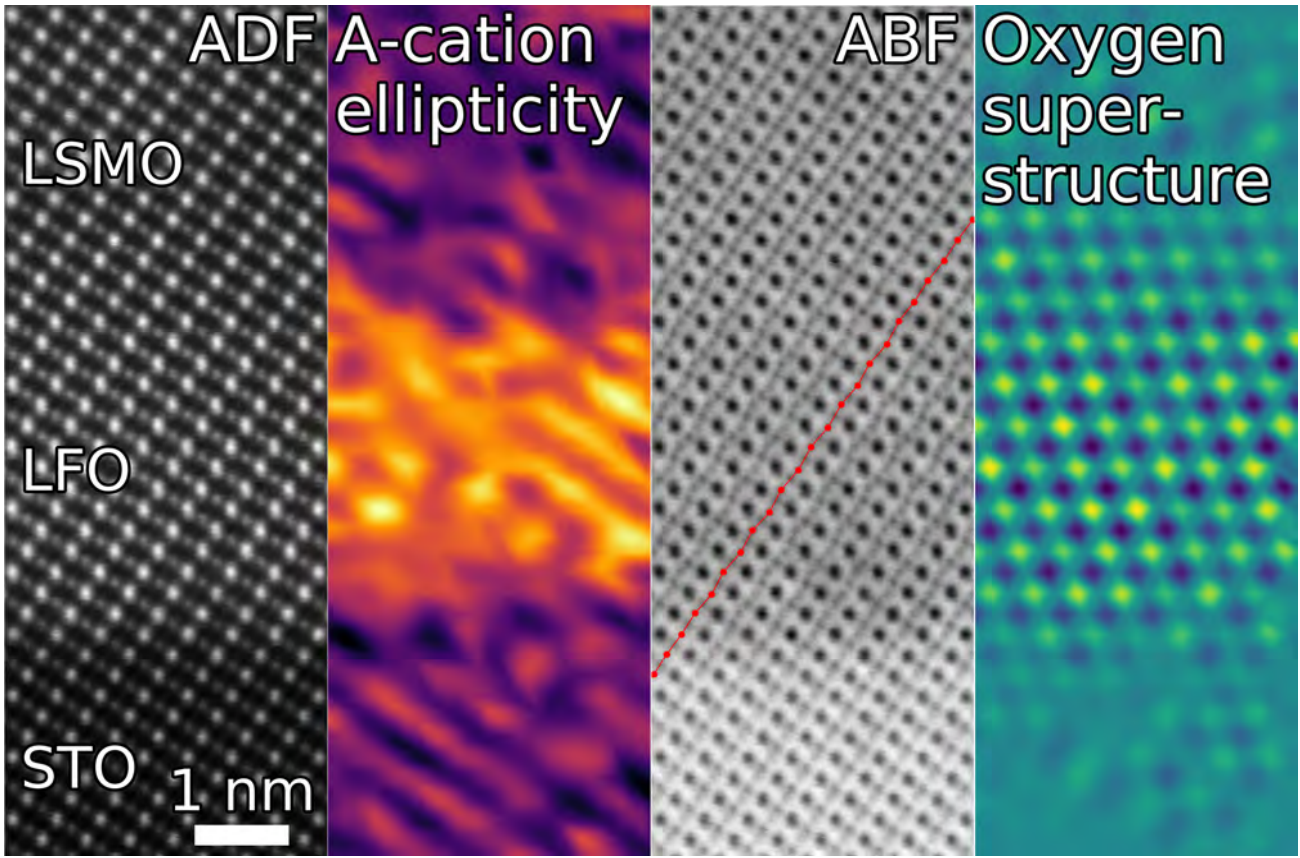

Figure 1: Atomic resolution STEM imaging and structural quantification of the bilayer structure using Atomap[18]: a) HAADF image, showing elliptical La columns in the $LaFeO_3$ layer; b) quantification of the A-cation ellipticity using 2D Gaussian fitting in Atomap; c) ABF image; d) quantification of O shifts away from the position in a primitive perovskite cell as a function of position, showing that this is mainly concentrated in the $LaFeO_3$ layer, and strongest at its centre.

In order to obtain information about the crystal structure parallel to the electron beam across the bilayer system full 4D STEM datasets were acquired. As shown in Figure 2a taken from the STO, these diffraction patterns contain a wealth of information including Kikuchi patters and HOLZ rings, the innermost of which is marked by the arrow. For comparison, a diffraction pattern from the LFO film is shown in Figure 2b and an extra inner HOLZ ring is clearly visible, at a radius which clearly demonstrates that there is symmetry breaking in the LFO film leading to a doubling of the size of the unit cell along the beam direction. Figure 2c presents a HAADF image calculated using the 4D dataset using all electrons scattered to > 106 mrad with the contrast dominated by the average atomic number of each layer, and clearly showing the interface between the films and the substrate. The HOLZ rings are quantified by centring and radially integrating each diffraction pattern (as is routinely performed in RDF analysis[38] and STEM fluctuation electron microscopy[39]), reducing the 4D dataset to 3D, using the pixStem software library[40], which is an extension of HyperSpy[41]. This gives the electron intensity as a function of scattering angle in each pixel as shown in Figure 2d. The background is removed by fitting a power law to the regions preceding and following the HOLZ peak and subtracting this to reveal the HOLZ peak itself.  This can then be fitted using Gaussian peak fitting, which then can be used to give parameters for total intensity, height, width and position.  Figure 2e is a plot of the total intensity in the inner Laue zone, which clearly shows the inner HOLZ ring is only generated in the LFO layer, and moreover that its strength extends across the LFO film region. Figure 2f shows a plot of three measures as a function of position from the LFO/STO interface, inner HOLZ ring intensity, A-site ellipticity (calculated from the data shown in Figure 1b), and oxygen shift from the undistorted perovskite position (calculated from the data shown in Figure 1d).  The three plots coincide very closely demonstrating that the A-site modulation, the period-doubling along the beam direction, and the octahedral tilting are all clearly coupled effects.  Therefore, any explanation of one effect must explain all three.

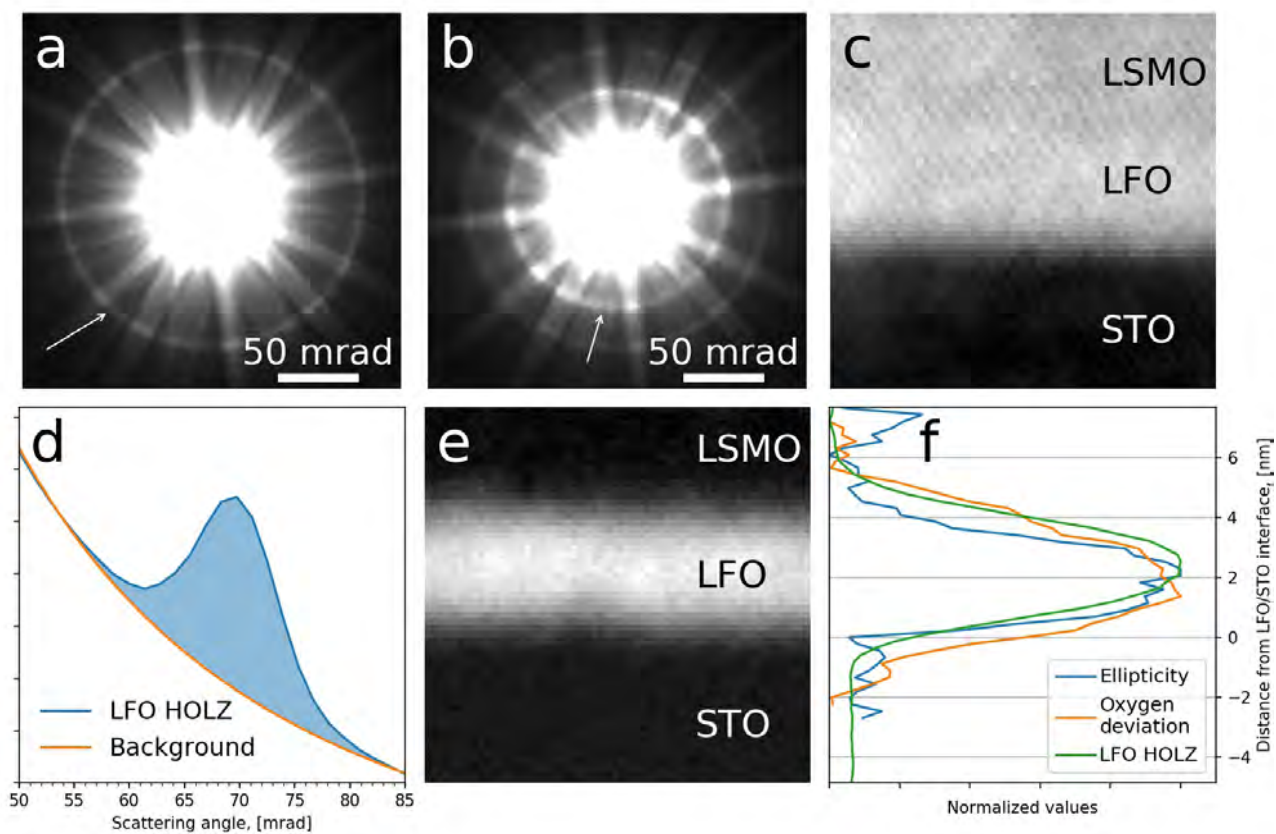

Figure 2: 4D STEM analysis of the LSMO/LFO/STO heterostructure along a <110> direction of the STO: a) STEM diffraction pattern from the STO substrate with an arrow indicating the outer HOLZ ring that appears in all three perovskite layers; b) STEM diffraction pattern from the centre of the LFO film with an arrow indicating the inner HOLZ ring that only appears in the LFO; c) HAADF image calculated from the 4D dataset with an inner angle of 106 mrad; d) Radial integration of the 4D dataset, yielding a 3-D dataset with electron counts as a function of scattering angle. Processing of the inner HOLZ ring, the background being a power law. (e) Intensity of the inner HOLZ ring. (f) Line profile of the ellipticity of the A-site columns, the oxygen shift from centrosymmetric position (see Figure 1), and the inner HOLZ ring intensity.

To more clearly understand the origins of the inner HOLZ ring, and its connection to the unit cell doubling of LFO, position averaged convergent beam electron diffraction (PACBED) STEM simulations were performed. Using the data shown in Figures 1 and 2, a suitable trial structure was built based on the bulk LFO structure consistent with this data and compatible with a STO-

(111) substrate, and was then relaxed using density functional theory (DFT) to form a triclinic P1 structure, rather than the orthorhombic *Pnma* structure seen in the bulk. This low symmetry arises from placing a nominally orthorhombic structure on a (111) face of the primitive structure, straining it slightly so that the orthorhombic symmetry is broken. This structure was the basis for PACBED [42] simulations using Dr Probe[43], where the amount of structural distortion was varied from the "bulk-like" fully distorted ($d$=1), to a non-distorted structure close to cubic ($d$=0) (full details of these structures are included in the Supplemental Information [44]). This distortion parameter, $d$, is used to scale the amount of oxygen octahedral tilt and the shifts of the La-atoms between the fully distorted value and zero. The resulting PACBED simulations of a fully distorted and non-distorted structure are shown in Figure 3a as opposite halves of a diffraction pattern. The fully distorted structure shows the same inner HOLZ ring as seen in the experimental data in Figure 2b, while no such ring is seen in the non-distorted structure. To quantify the intensity of the HOLZ rings, the same radial integration is done on these simulated diffraction patterns as for the experimental patterns, presented in Figure 3b. This shows that a more intense inner HOLZ ring is due to a higher amount of distortion. To determine which of the atom movements contribute the most to the inner HOLZ ring intensity, the simulations were also done on a distorted structure where only the La atoms were displaced, but the oxygen atoms were left in an unshifted position (not shown), Fe atoms were not moved because they do not appear to modulate and the octahedra appear to tilt around centres fixed by the Fe atoms. Although such a distortion is unrealistic, it is valuable to understand the origin of the HOLZ ring. Such simulations showed that most of the HOLZ ring intensity persists in this case too, and therefore it is clear that most of the intensity in the HOLZ ring originates from the La columns, while the oxygen displacements only contribute a smaller amount. This is expected, due to the much larger electron scattering cross section of La. Nevertheless, the modulation of the La atoms along this crystal direction and rotation of the oxygen octahedra about the same direction are clearly coupled effects in this crystal structure, and not really separable, and thus, as demonstrated in Fig 2f, the inner Laue zone intensity and the octahedral tilting correlate closely, meaning that measuring the Laue zone intensity is a good proxy for measuring the octahedral tilting directly from quantification of oxygen atom positions, at least for this structure.

The fully distorted modified LFO structure as reconstructed by a combination of the experimental observations and the DFT relaxation is shown in Figure 3c, in the same projection as the STEM image in Figure 1. The ellipticity and oxygen distortions are similar to those experimentally observed in Figure 1, and the degree and pattern of movements of La atoms along the viewing direction accounts for the Laue zone intensity plotted in Figure 2 and simulated in Figure 3a and 3b. Hence, the PACBED model of a cell slightly distorted from orthorhombic to triclinic describes the data well. It should be noted, however, that although this new structure for LFO is derived from the bulk orthorhombic structure by compressive strain from growth on a (111) perovskite surface, it displays some significant changes from the bulk crystal structure. Specifically, the modulation of the A-sites also happens in bulk LFO, but along a different axis to the octahedral tilting axis, whereas these two axes are coincident in the (111) growth-stabilised structure. More information about this strain-induced structure is given in the Supplemental Information [44].

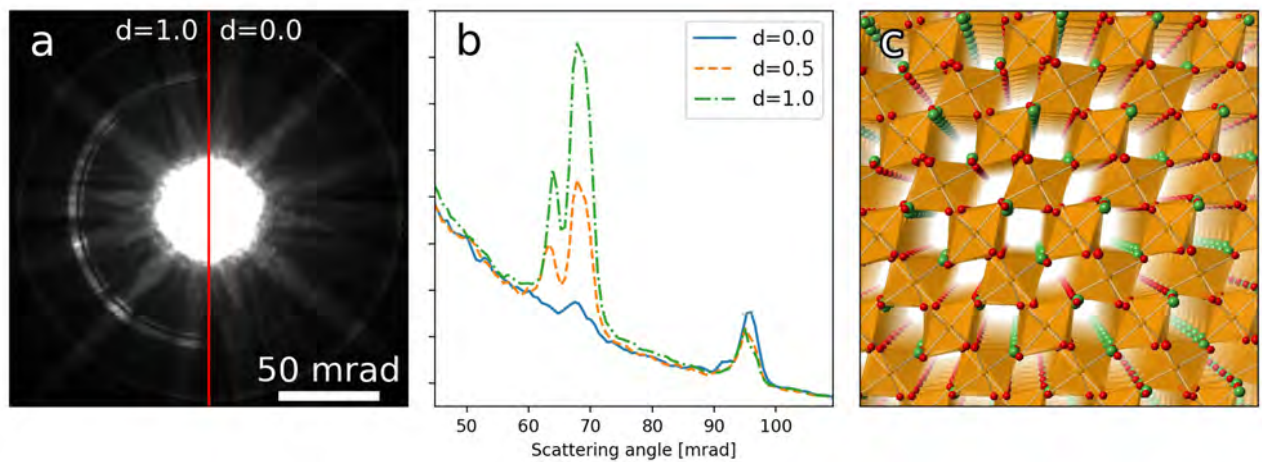

Figure 3: Results from PACBED STEM simulations. (a) Simulated diffraction image from fully distorted, and not distorted modified LFO. The fully distorted showing the extra HOLZ ring. (b) Radial integration of the simulated diffraction patterns with different amounts of distortion. (c) Pseudo-3D visualization of the modified LFO structure model, with La (green), Fe (orange) and O (red).

What we observe in the sandwiched LFO is that the tilt transition in the octahedra and the accompanied cation modulation is quenched close to the interfaces to the STO and to the LSMO – and the structures seen in each layer are shown schematically in Figure 4. It is expected that the tilting in the LFO should be suppressed at the interface with the STO since this substrate has a simple cubic structure with no octahedral tilting (i.e. $a^0a^0a^0$ in Glazer notation). Such an effect of tilt suppression near interfaces has been seen previously in growth of thin films on STO or other cubic perovskites [13,14,16,37], or in $BiFeO_3$ on a very thin (~ 5nm) LSMO layer on STO, where the LSMO was transformed by strain to an untilted tetragonal structure [11]. For the interface with LSMO, the quenching of the LFO tilt pattern likely occurs because the two materials have incompatible tilt patterns, which is illustrated in more detail in the Supplemental Materials [44] in Figures S2 and S3. LSMO in bulk is well known to have an $a^-a^-a^-$ tilt pattern in a rhombohedral cell, resulting in antiphase rotations about the film normal and the pseudocubic <100> axes, but no noticeable tilting about the <110> viewing direction using in this work. The refined LFO cell shown shows a $a^-b^+c^-$ tilt pattern and shows in-phase tilting about the pseudocubic [010] axis, as well as about the viewing direction. Thus, this tilt-incompatibility could well result in the need to have zero tilt about the viewing direction used in Figures 1 and 2 at the LFO-LSMO interface, and therefore in suppression of the tilting and associated A-site modulation in the upper part of the LFO film.

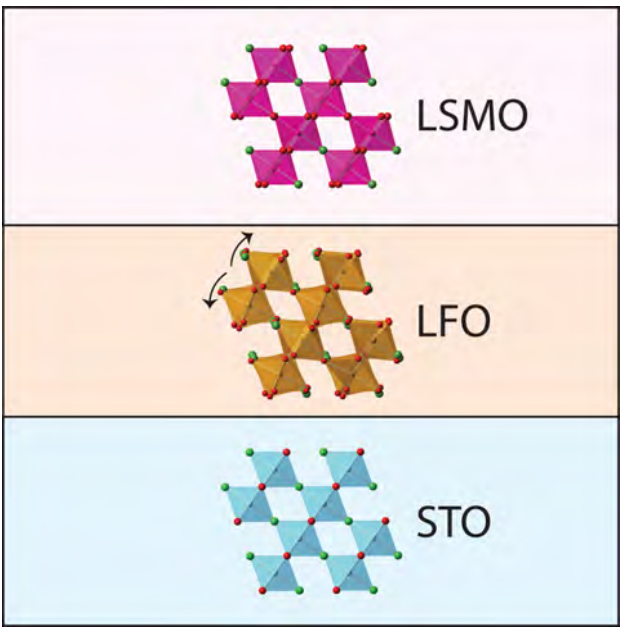

Figure 4: Schematic diagram of the tilting patterns along the $[1\bar{1}0]_{STO}$ direction for the three structures, provided the LSMO is oriented with $[0001]_{LSMO}//[111]_{STO}$. The STO and LSMO layers show no rotations of octahedra about this direction, whereas the sandwiched LFO layer has a structure of coupled rotations of octahedra about this axis, which would have to be quenched at the interfaces to match the untilted structures to either side.

Imaging of several regions of the LFO thin film revealed two different domain configurations. Some show all the features discussed above of octahedral tilting, elliptical La columns and the strong inner Laue zone. Others show none of these features, and a comparison of two STEM diffraction patterns is shown in Figure 5 to illustrate this point. This is consistent with expectations for growing a nominally orthorhombic structure on a (111) facet: there should be at least 3 growth orientations resulting in a domain pattern as shown. If, as in our refined structure, only two of the in-plane <110>$_{primitive}$ directions supports the zig-zag period-doubled A-site columns and the octahedral tilting perpendicular to that direction, then only two domains in three in the film will produce the extra Laue zone, as shown.

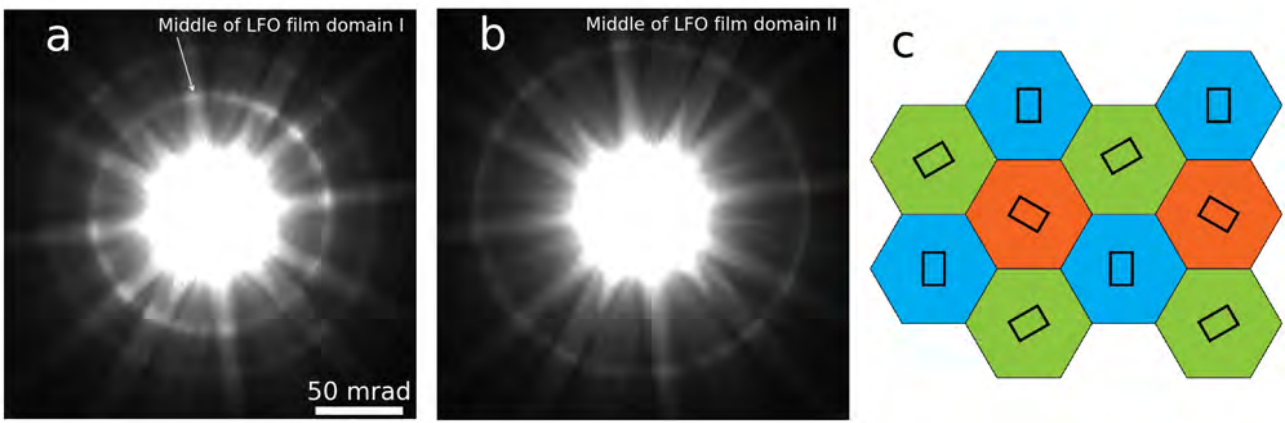

Figure 5: STEM diffraction imaging from a) domain I, and b) domain II; c) a schematic diagram of how 3-distinct domain orientations can form on the (111) surface (viewed along the surface normal), and in this case, this means 2 domains in 3 would be expected to show the period doubling, irrespective of which <110>$_{primitive}$ direction is used for imaging.

Considering more broadly the possible uses of the Laue zone ring imaging technique introduced here, there are a number of areas where it could be advantageous. One of the advantages of the technique is that this imaging is rather easy to perform compared to the careful HAADF imaging needed to perform the ellipticity determination on individual columns, provided a suitable detector is present for recording the diffraction patterns.  Moreover, it is more straightforward than high quality ABF imaging needed to directly measure the octahedral tilting, which always requires a great deal of operator skill, as well as a perfect sample.  Consequently, for mapping larger areas of specimens where the periodicity along the third dimension may be changing as a consequence of octahedral tilting or some other structural transformation, this would be a very useful technique.  That could include strain-induced transformations in thin films or heterostructures as in the present work, but could also include changes due to domain structure formation in ferromagnets, ferroelectrics, ferroelastics, or martensitic alloys.

Moreover, we note that if the angular centre of the Laue zone ring were plotted as a function of position, this could be used to plot lattice parameter changes with position, allowing, for example to examine strain relaxation around interfaces and strained thin films and heterostructures.

If this kind of Laue-zone STEM imaging could be performed using an aberration-corrected Ångström-sized probe, then it would be possible to perform atomic resolution studies of the variation in periodicity between different sites in a crystal.  In principle, this has been done previously by Huang *et al*. in sodium cobaltate [25], although the more primitive detector setup meant that the contrast in that study was a mixture of standard high-angle incoherent scattering and coherent HOLZ scattering, which made that work less quantitative than is now possible using pixelated direct electron counting detectors as shown here.  Beyond that, the actual HOLZ ring radius should vary depending which column the Bloch wave is centred on [45], which can result in split HOLZ rings in crystals because different Bloch waves run along regions at different potentials. It may be possible therefore with atomic resolution Laue zone imaging to resolve these differences, and possibly even make direct real-space relative measurements of the location and relative potential of different Bloch waves.

### IV.     CONCLUSIONS

Using a 4D STEM approach we can map the periodicity and the magnitude of atomic movements along the beam direction by imaging the intensity of scattering into specific higher order Laue zone rings.  This adds information about the third dimension to STEM characterisation of crystals, including cell doubling from octahedral tilting and associated cation movements, without needing to make multiple lamellae or use high tilt tomography.  In combination with atomic resolution imaging of the same lamella, this allows a lot of detail about the local 3D "crystal" structure of a perovskite oxide thin film system to be found from one orientation alone, in this case allowing an approximate atomic model to be developed that could be further refined using DFT. In this work, we showed that an additional inner HOLZ ring appears in a $LaFeO_3$ layer sandwiched between a $SrTiO_3$ substrate and a $La_{0.7}Sr_{0.3}MnO_3$ top layer, and that this HOLZ ring is strongest at the centre of the layer. This correlates with the strength of oxygen octahedral tilting in this layer and with a coupled modulation of La positions, which is visible as an elliptical cross section for the La columns in the $LaFeO_3$ layer in HAADF images. It is shown by image simulation that the La modulation accounts for the majority of the Laue zone intensity. In this case, the fact that the La-modulation and the octahedral tilting are coupled allows us to use the Laue zone measurements as a simple and straightforward way to map octahedral tilting strength over large areas of a thin film. This allowed us to reconstruct the 3D periodicity of the film on a cell by cell basis through the film, and revealed that the local ordering in this $LaFeO_3$ displays noticeable differences to that in the bulk orthorhombic structure. On the basis of these observations we were able to construct an approximate structure which was relaxed using DFT methods to a triclinic cell. It was also shown that the cell tilting and A-site modulation is suppressed at both the interface to the $SrTiO_3$ and to the $La_{0.7}Sr_{0.3}MnO_3$, presumably because no tilting is present in the first, and because an incompatible tilt pattern is formed in the latter.

It is also shown that this 4D Laue STEM imaging could have a number of other applications including the study of domain structures in ordered materials, ferroelectrics and similar, and atomic resolution studies of stacking sequences in more complex structures.

## ACKNOWLEDGEMENTS

We are grateful to Prof. P.D. Nellist for helpful discussions on this work. This work was funded by the Engineering and Physical Sciences Research Council via the grant "Fast Pixel Detectors: a paradigm shift in STEM imaging" (EP/M009963/1). TT acknowledges Research Council of Norway grant 231290.